\documentclass[twocolumn,epjc3]{svjour3}
\RequirePackage{graphicx}
\RequirePackage{amsmath}
\RequirePackage{amssymb}
\RequirePackage{amsfonts}
\RequirePackage{units}
\usepackage{color}
\usepackage{enumerate}
\newcommand{\pt}{p_\perp}
\newcommand{\kt}{k_\perp}
\renewcommand{\d}{\text{d}}
\newcommand{\jewel}{\textsc{Jewel}\ }

\journalname{Eur. Phys. J. C}

\begin{document}    

\title{{\hfill \normalsize \rm{CERN-PH-TH-2015-310, MCnet-15-31}\\ \medskip \\}
	Origins of the di-jet asymmetry in heavy ion collisions}


\author{Jos\'e Guilherme Milhano\thanksref{e1,addr1,addr2}
        \and
        Korinna Christine Zapp\thanksref{e2,addr1,addr2} 
}

\thankstext{e1}{e-mail: guilherme.milhano@tecnico.ulisboa}
\thankstext{e2}{e-mail: korinna.zapp@cern.ch}


\institute{CENTRA, Instituto Superior T\'ecnico, Universidade de Lisboa, Av. Rovisco Pais, P-1049-001 Lisboa, Portugal \label{addr1}
           \and
           Physics Department, Theory Unit, CERN, CH-1211 Gen\`eve 23, Switzerland \label{addr2}
}

\date{Received: date / Accepted: date}

\maketitle

\begin{abstract}

The di-jet asymmetry --- the measure of the momentum imbalance in a di-jet system --- is a key jet quenching observable. Using the event generator \jewel we show that the di-jet asymmetry is dominated by fluctuations both in proton-proton and in heavy ion collisions. We discuss how in proton-proton collisions the asymmetry is generated through recoil and out-of-cone radiation. In heavy ion collisions two additional sources can contribute to the asymmetry, namely energy loss fluctuations and differences in path length. The latter is shown to be a sub-leading effect. We discuss the implications of our results for the interpretation of this observable.

\end{abstract}

\section{Introduction}
\label{intro}
The ability to systematically reconstruct jets above the large and fluctuating background present in ultra-relati\-vistic heavy ion collisions has opened up a versatile path to study the properties of Quark Gluon Plasma (QGP). 
Jets are sensitive, through the wide range of scales involved in their development, to a variety of properties of the expanding QGP they traverse. 
Unlike measurements that involve hadrons (e.g. single hadron suppression), jet observables are mostly immune to the uncertainties arising from the ill-understood physics of hadronization.

The extensive use of jets in both hadron and lepton collisions is grounded on solid theoretical understanding. Both the jet production and jet evolution giving rise to the characteristic jet structure are calculable in perturbation theory and are encoded in Monte Carlo event generators. This is in contrast with the present situation in heavy ions where, albeit very important theoretical developments in the last few years (for a recent review see \cite{Mehtar-Tani:2013pia}), the dynamical details of jet-medium interactions remain partly ununderstood.  

Although current Monte Carlo implementations of jet dynamics in the presence of a medium are necessarily incomplete, they can be used meaningfully in a variety of studies. Ultimately, the endowment of jets with full probing potential requires  the dependence of a given jet observable on specific medium properties to be clearly identified. By considering an event generator --- \textsc{Jewel}~\cite{Zapp:2012ak,Zapp:2013vla} --- that has been validated for a wide set of observables and a specific observable, we illustrate a generic strategy for achieving such identification. 

We carry out a detailed analysis of what drives the enhancement of di-jet energy imbalance in heavy ion collisions relative to the proton-proton case. In doing so, we attempt to qualify common assumptions made in the literature. Di-jet asymmetry carries the historical weight of having been the first observable to be measured for fully reconstructed jets in heavy ion collisions~\cite{Aad:2010bu} and of having triggered nearly immediate insight on the underlying dynamics at play~\cite{CasalderreySolana:2010eh,Qin:2010mn}. Since then more differential measurements~\cite{Chatrchyan:2012nia} and attempts to observe a di-jet asymmetry at RHIC~\cite{Kauder:2015vvr} have been carried out.

The di-jet asymmetry
\begin{equation}
\label{eq:aj_def}
A_J = \frac{p_{\perp, 1} - p_{\perp, 2}}{p_{\perp, 1} + p_{\perp, 2}}\, ,
\end{equation}
measures the imbalance between the transverse momenta $p_{\perp, 1}$ of the leading jet and $p_{\perp, 2}$ of the sub-leading jet in a di-jet pair. 

As with any observable, a number of confounding factors are necessarily at play. While naively one would expect the difference in the (matter weighted) path lengths of the jets in the pair to be a leading factor to the generation of the observed increase in asymmetry in heavy ion collisions, our study strongly suggests otherwise. Instead, we find that the asymmetry enhancement results from the aggregate effect of `vacuum-like' and medium induced fluctuations.

The note is organized as follows. In sec.~\ref{sec:setup} we describe the setup underlying our study, highlighting the salient features of \textsc{Jewel} and providing details for the Monte Carlo di-jet sample we use.  Sec.~\ref{sec:results} presents the results of the study, establishing the origins of the di-jet asymmetry in both proton-proton and heavy-ion collisions. Finally, in sec.~\ref{sec:summary} we summarize our main findings and discuss the wider lessons learnt from our study.

\section{Setup}
\label{sec:setup}
\subsection{Jet evolution in \textsc{Jewel}}

\textsc{Jewel}~\cite{Zapp:2012ak,Zapp:2013vla} is a Monte Carlo event generator for jets in proton-proton and heavy ion collisions. Jet production, QCD scale evolution and re-scattering of jets in a background medium are described in a common perturbative framework. Both the initial hard process giving rise to hard partons and re-scattering are described by infra-red continued leading order $2\to 2$ matrix elements. Radiative corrections to both kinds of processes are generated by the same parton shower, which is thus responsible for jet evolution and medium induced radiation. All emissions have a finite formation time. In cases where there are competing sources of radiation, i.e.\ the initial jet production and a re-scattering, the emission with the shorter formation time is realised. This has the important consequence that a re-scattering, which is typically soft or semi-hard, cannot perturb the evolution of a highly virtual parton. When the formation times of emissions associated to several re-scatterings overlap, these act coherently to emit a single gluon (this is the well-known LPM effect, for a discussion of how this can be realised in an event generator see~\cite{Zapp:2011ya}). In proton-proton collisions \textsc{Jewel}'s parton shower reduces to a standard vacuum parton shower.

Hard jet production matrix elements and the corresponding initial state parton showers as well as hadronisation and hadron decays are generated by \textsc{Pythia}\,6.4 \cite{Sjostrand:2006za} using the \textsc{Eps09} nuclear PDF sets~\cite{Eskola:2009uj} together with \textsc{Cteq6LL}~\cite{Pumplin:2002vw}, both provided by \textsc{Lhapdf}~\cite{Whalley:2005nh}.

\begin{figure}
	\includegraphics[width=\linewidth]{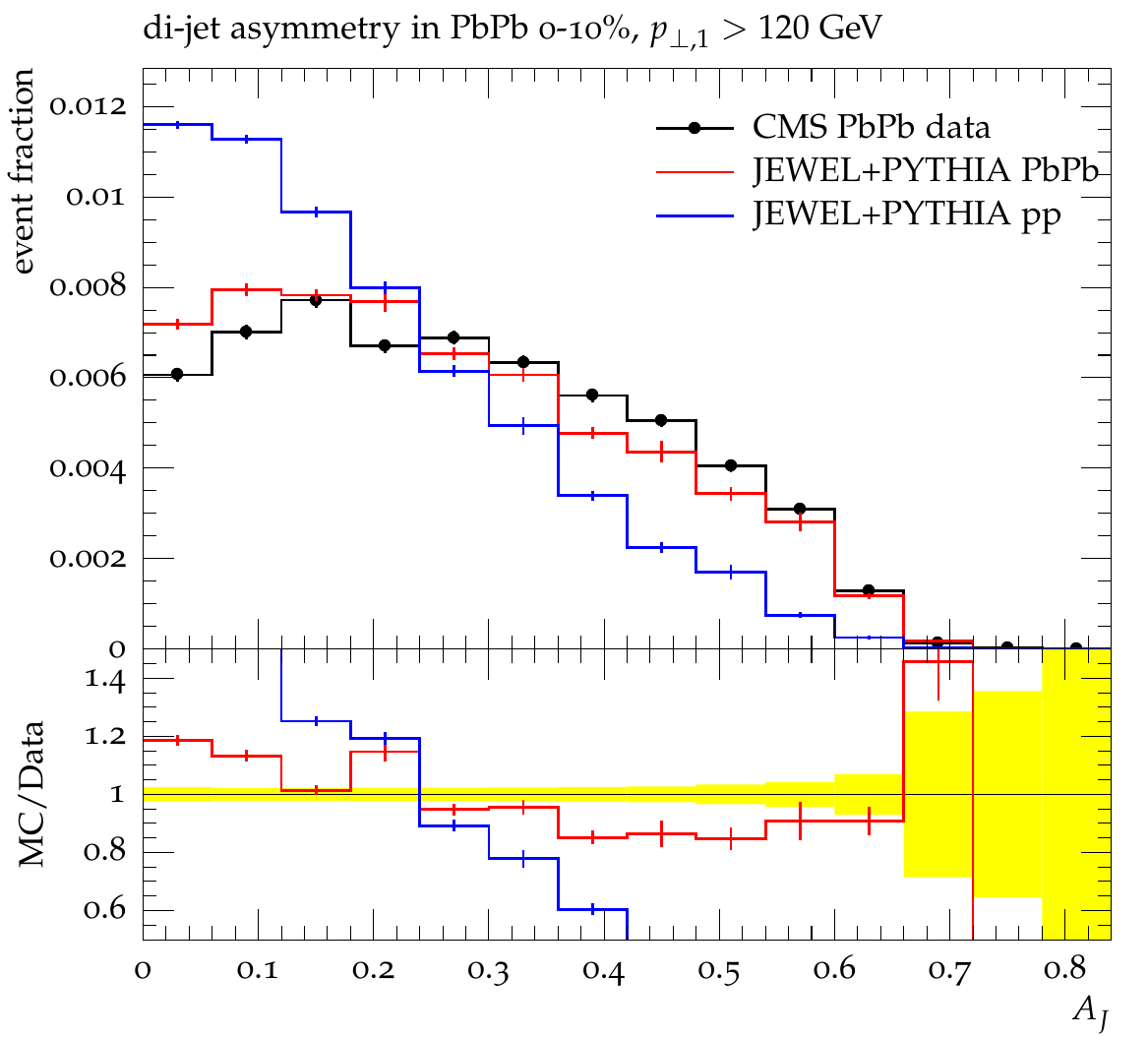}
	\caption{\textsc{Jewel} results compared to CMS data~\cite{Chatrchyan:2012nia} for the di-jet asymmetry in central PbPb collisions at $\sqrt{s_\text{NN}} = \unit[2.76]{TeV}$ for anti-$k_\perp$ jets with $R=0.3$ and $|\eta|<2$. Di-jet cuts are $p_{\perp,1} > \unit[120]{GeV}$ and $p_{\perp,2} > \unit[30]{GeV}$ and $\Delta \phi_{12} > 2\pi/3$. The Monte Carlo events are smeared with the parametrised resolution from~\cite{Chatrchyan:2012gt}, as the data are not unfolded for jet energy resolution.}
	\label{fig::ajcms}
\end{figure}

For sufficiently hard observables \textsc{Jewel} describes a large variety of data reasonably well. As an example figure~\ref{fig::ajcms} shows the di-jet asymmetry as measured by CMS (comparisons to other measurements can be found in~\cite{Zapp:2013vla,Zapp:2013zya}).

\subsection{The generated di-jet sample}

This study is based on a di-jet sample generated with \textsc{Jewel}\,2.0.2 in the set-up discussed in~\cite{Zapp:2013vla} at a nucleon-nucleon centre of mass energy $\sqrt{s_\text{NN}} = \unit[2.76]{TeV}$. The simple, parametrised background described in detail in \cite{Zapp:2013zya} is used, as we don't have evidence that running with a full hydrodynamic background leads to significant effects. The initial time and temperature are taken as $\tau_\text{i} = \unit[0.6]{fm}$ and $T_\text{i} = \unit[485]{MeV}$ as in~\cite{Shen:2012vn}, the critical temperature is $T_\text{c} = \unit[170]{MeV}$. As this study is concerned with effects best discussed in azimuthally symmetric events, we generate only the most central events with vanishing impact parameter $b=0$. A matching sample is generated for proton-proton collisions.

Jets are reconstructed with the anti-$k_\perp$ algorithm \cite{Cacciari:2008gp} provided by the \textsc{FastJet} package~\cite{Cacciari:2011ma} with a radius parameter of $R=0.4$ within $|\eta| < 2$. Recoils are not included, so no background subtraction is necessary (cf. discussion in~\cite{Zapp:2013vla}). The di-jet cuts are $p_{\perp,1} > \unit[100]{GeV}$ for the leading and $p_{\perp,2} > \unit[20]{GeV}$ for the sub-leading jet and an azimuthal separation $\Delta \phi_{12} > \pi/2$ between the two jets.

\smallskip

For a part of the discussion it is necessary to match a jet to the initial parton from the matrix element. This cannot be done in di-jet events, therefore we generated a sample of $\gamma$-quark jet events (for this an unpublished extension of \textsc{Jewel}\,2.0.2 was used). These events are only used to make a qualitative observation for which the flavour composition of the jets is irrelevant. To faciliate the parton-jet matching the initial state parton shower was disabled. The medium set-up in this sample is the same as for the di-jets. 

For the analysis all photons with $\pt > \unit[5]{GeV}$ are removed from the event before the jets are reconstructed, again with the anti-$\kt$ algorithm and $R=0.4$. In addition to the jet cuts, which are $|\eta| < 5$ and $\pt > \unit[20]{GeV}$, the initial parton is required to be within $|\eta| < 2.5$ and pass the same $\pt$ cut to avoid cases where no jet is reconstructed because there is no initial parton that could give rise to a jet in the required phase space. 

\smallskip

Monte Carlo events are analysed and histograms plotted with Rivet~\cite{Buckley:2010ar}.

\section{Origins of the di-jet asymmetry}
\label{sec:results}

In proton-proton collisions, the di-jet asymmetry is induced entirely by fluctuations in the fragmentation pattern. A pair of hard partons produced by a lowest order $2\to 2$ scattering process (described by a matrix element) cannot have an asymmetry. The large and well-known radiative corrections in the form of extra emissions, however, induce an asymmetry. In Monte Carlo event generators these corrections are generated to leading logarithmic accuracy by the parton shower\footnote{Multi-jet configurations are often better described by multi-leg matrix elements matched to a parton shower. But also a multi-leg matrix element can be related to a $2\to 2$ core process by clustering backwards (e.g.\ by running the parton shower backwards) until a $2\to 2$ configuration is reached. We shall therefore regard all extra emissions as corrections to a $2\to 2$ core process, irrespective of whether they were generated with a matrix element or a parton shower, and whether or not they give rise to additional jets.}.

\smallskip

In heavy ion collisions, two additional sources contribute to the asymmetry: the difference in path-length between the leading and the sub-leading jet, and energy loss fluctuations. 
In the following we shall argue that the di-jet asymmetry in heavy ion collisions is dominated by fluctuations and that the effect of path-length difference is small. This statement does not imply that the di-jet asymmetry is independent of path-length. A clear dependence on average path-length can be ascertained by studying the variation of $A_J$ with the angle of the di-jet system relative to the reaction plane in non-central events. However, such dependence is not relevant for the arguments put forward in this study where, for simplicity, we restrict the discussion to azimuthally symmetric (i.e. the most central) events. 
Importantly, we show that not only energy loss fluctuations, but also hard fluctuations that cannot be attributed to jet-medium interaction play an important role in the increase of asymmetry in heavy ion collisions.

\subsection{Effect of path-length difference}

\begin{figure}
\includegraphics[width=\linewidth]{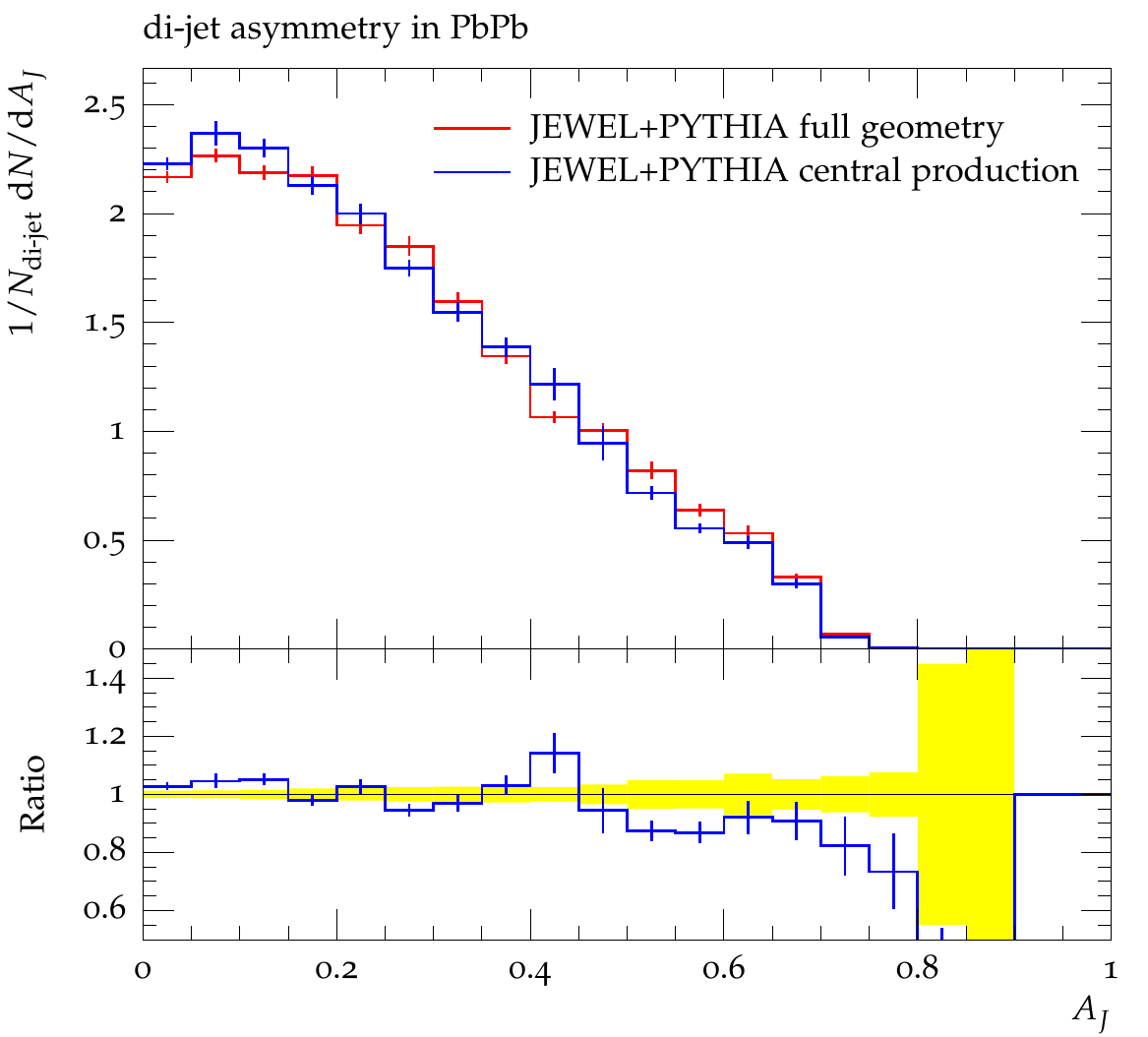}
\caption{Di-jet asymmetry $A_J$ in central ($b=0$) Pb+Pb events in a scenario where the di-jet production points are distributed according to the Glauber model ('full geometry') compared to a scenario where all jets are produced at the centre of the collision ('central production').}
\label{fig::killerplot}
\end{figure}

To clarify the effect of path-length difference between leading and sub-leading jets,  we start by comparing two scenarios: (i) `full geometry' where di-jet production points are realistically distributed according to the Glauber model, and (ii) `central production'  in which all di-jets  are produced in the centre of the collision. In the latter scenario the path-lengths are obviously the same. 
If in the sample with distributed production points  a strong bias for the leading jet to have the smaller path-length was present and such difference was driving the asymmetry, then the di-jet asymmetry should be significantly larger in this scenario than in the `central production' case where all path-lengths are the same. Figure~\ref{fig::killerplot} shows clearly that this is not the case. The difference between the asymmetry computed in the two scenarios is small.
This provides clear evidence that fluctuations, rather than  systematic path-length differences, are most relevant in  building up the asymmetry.


In \textsc{Jewel}, and arguably in general, jet-medium interaction depends on the amount of medium traversed by the jet. The relevant path-length that accounts for the evolving medium density profile is the density weight\-ed path-length given by 
\begin{equation}
L_n = 2 \frac{\int\!\d \tau\, \tau n(\mathbf{r}(\tau),\tau)}{\int \!\d \tau\, n(\mathbf{r}(\tau),\tau)} \,,
\end{equation}
where $\tau = \sqrt{t^2-z^2}$ is the proper-time and $n(\mathbf{r}(\tau),\tau))$ is the position and time dependent density of medium scattering centres. As we consider a boost invariant medium, $L_n$ is rapidity independent.

\begin{figure}
\includegraphics[width=\linewidth]{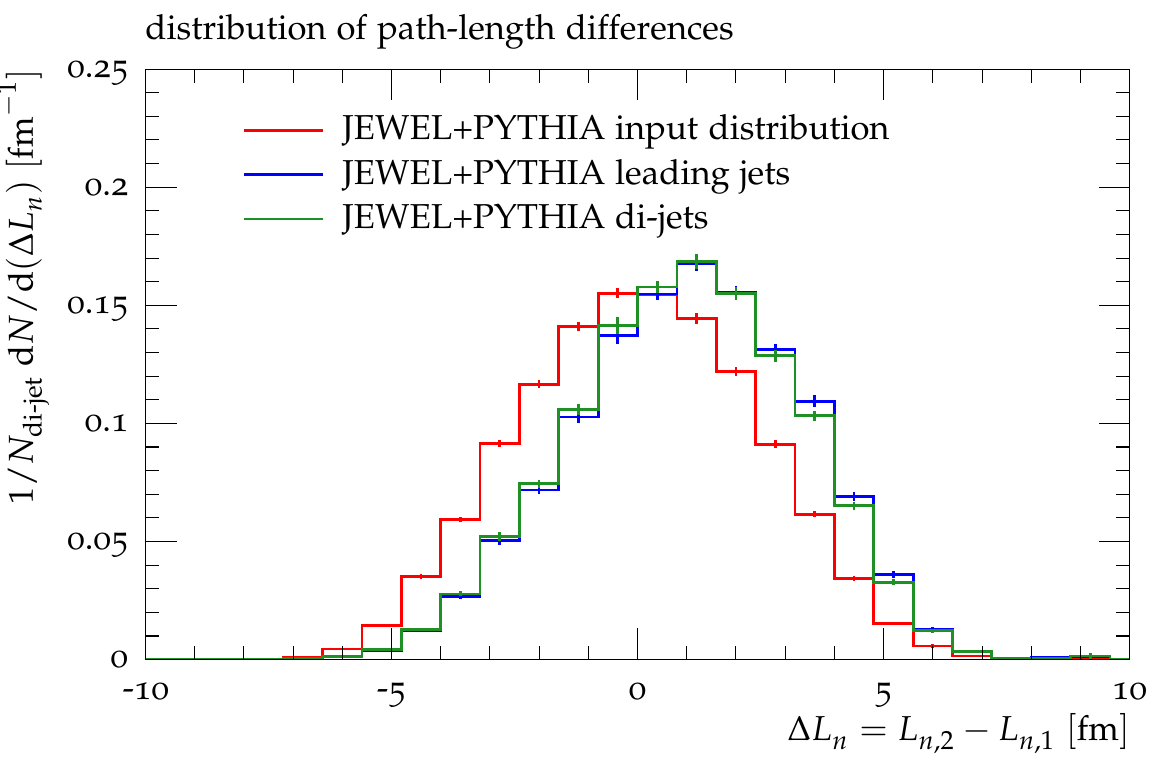}
\caption{Comparison of differences in path-length between leading and sub-leading jet when no jet cuts are placed (red), when only one jet passing the $\pt$ cut for the leading jet is required (blue) and when a di-jet system is required (green).}
\label{fig::deltaln}
\end{figure}

Figure~\ref{fig::deltaln} shows the distribution of the path-length difference ($\Delta L_n = L_{n,2} - L_{n,1}$) between the sub-leading and leading jet in  di-jet events, together with analogous distributions obtained in single-inclusive jet events and  without any jet cuts.
The path-lengths for the leading jet $L_{n,1}$ and sub-leading jet $L_{n,2}$ in each di-jet event are computed  from the di-jet production point and the direction of each of the reconstructed jets in the pair.
For single-inclusive jet events, the jet is required to pass the same leading jet $\pt$ cut as in di-jet events and the sub-leading jet, which is not reconstructed, is assumed exactly back-to-back (the azimuthal angle between the two jets is $\Delta \phi = \pi$).
The distribution in the case where no jet cuts are imposed simply reflects the Glauber distribution of production points. Here, the angles and transverse momenta of the out-going partons of the matrix element are used to evaluate the path-lengths. 

\begin{figure}
\includegraphics[width=\linewidth]{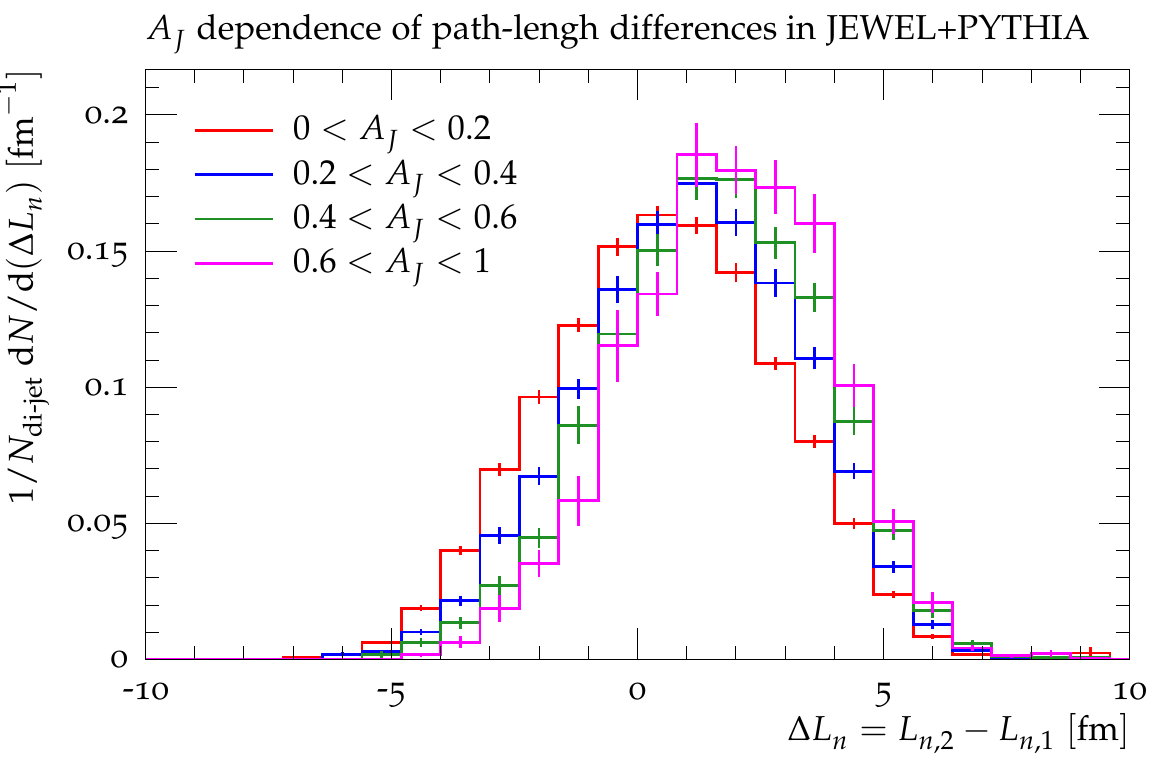}
\caption{Path length between leading and sub-leading jet in di-jet events for different di-jet asymmetries.}
\label{fig::deltaln-binned}
\end{figure}

The distribution without jet cuts is symmetric around zero, while both the di-jet and single-inclusive jet cases show a shift towards positive $\Delta L_n$. This shift, favouring somewhat smaller path-lengths for the leading jet, is a consequence of the $\pt$ cut imposed on the leading jet\footnote{The near coincidence of the distributions for the di-jet and single-inclusive jet cases results from the very asymmetric $\pt$ cuts ($p_{\perp,1} > \unit[100]{GeV}$ and $p_{\perp,2} > \unit[20]{GeV}$) that are imposed.}. This is not, however, a large effect. In fact, in \unit[34]{\%} of the di-jet systems the leading jet has the longer path-length. Such configurations are only possible in the presence of sizeable vacuum and/or medium energy loss fluctuations. As figure~\ref{fig::deltaln-binned} shows, there is a mild correlation between the path-length difference and the di-jet asymmetry (the mean path-length difference increases from $\langle \Delta L_n \rangle =0.56$ in the most symmetric to $\langle \Delta L_n \rangle = 1.86$ in the most asymmetric bin). This shift is still small compared to the width of the distribution, which is a measure for the importance of fluctuations.

The path-length of a jet produced in the centre is \unit[4]{fm}, while in the scenario with distributed production points the average path-length is \unit[3.74]{fm}. Therefore, there is room for a small effect due to path-length differences in figure~\ref{fig::killerplot}, but it cannot be large.

\subsection{Di-jet asymmetry in p+p}

\begin{figure}
\includegraphics[width=\linewidth]{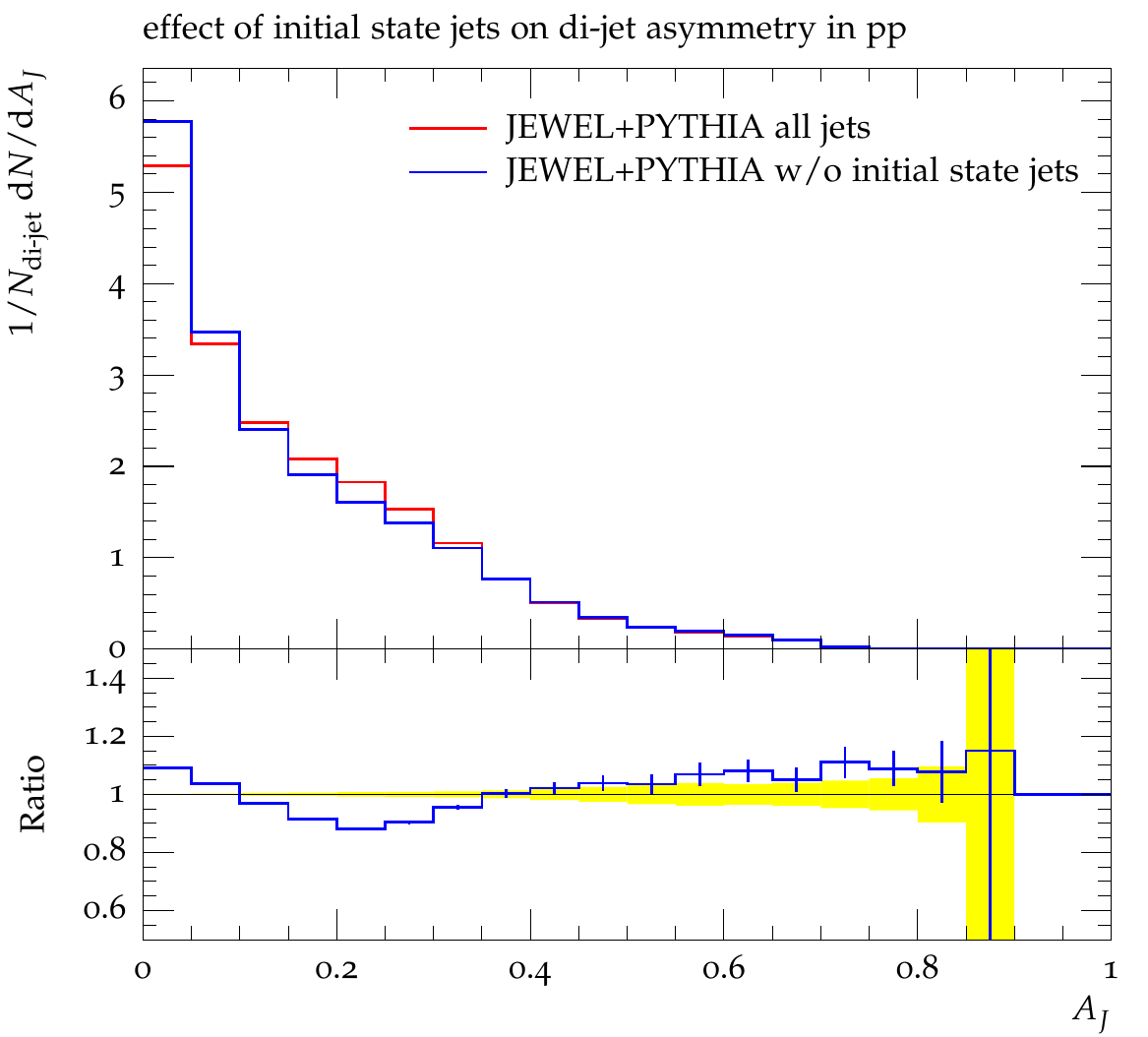}
\caption{Di-jet asymmetry in p+p in all events (red) and events where both jets are classified as coming from the final state (blue).}
\label{fig::ajvac}
\end{figure}

Before returning to the discussion of the mechanisms driving the increase of di-jet asymmetry in heavy-ion collisions, we now discuss the di-jet asymmetry in the vacuum case.

The di-jet sample inevitably suffers from contamination from initial state radiation as the two hardest jets in an event are not necessarily the result of the final state of the matrix element. One, or both, of them can originate from initial state emissions. These configurations are not relevant for our study and should be excluded from the sample. Since an unambiguous assignment of a jet to the initial or final state is not possible, we have  implemented an approximate procedure that compares the transverse momenta of the two reconstructed jets with those of the outgoing partons of the matrix element.  
We assume that the harder jet corresponds to the harder parton. If the transverse momentum of one of the jets is more than \unit[10]{\%} larger than that of the matching parton then this jet is classified as coming from the initial state and the event is rejected. As expected, it is far more likely that the sub-leading jet is classified as being due to an initial state emission. The procedure works reasonably well in practice and discards \unit[13]{\%} of the events. Figure~\ref{fig::ajvac} shows the comparison of the di-jet asymmetry obtained before and after rejection of events containing initial state jets. The small difference between the distributions ensures that the contamination from initial state jets will not compromise any of the conclusions of this study. In the remaining dijet events no matching between matrix element outgoing partons and reconstructed jets is carried out.

\smallskip

\begin{figure}
\includegraphics[width=\linewidth]{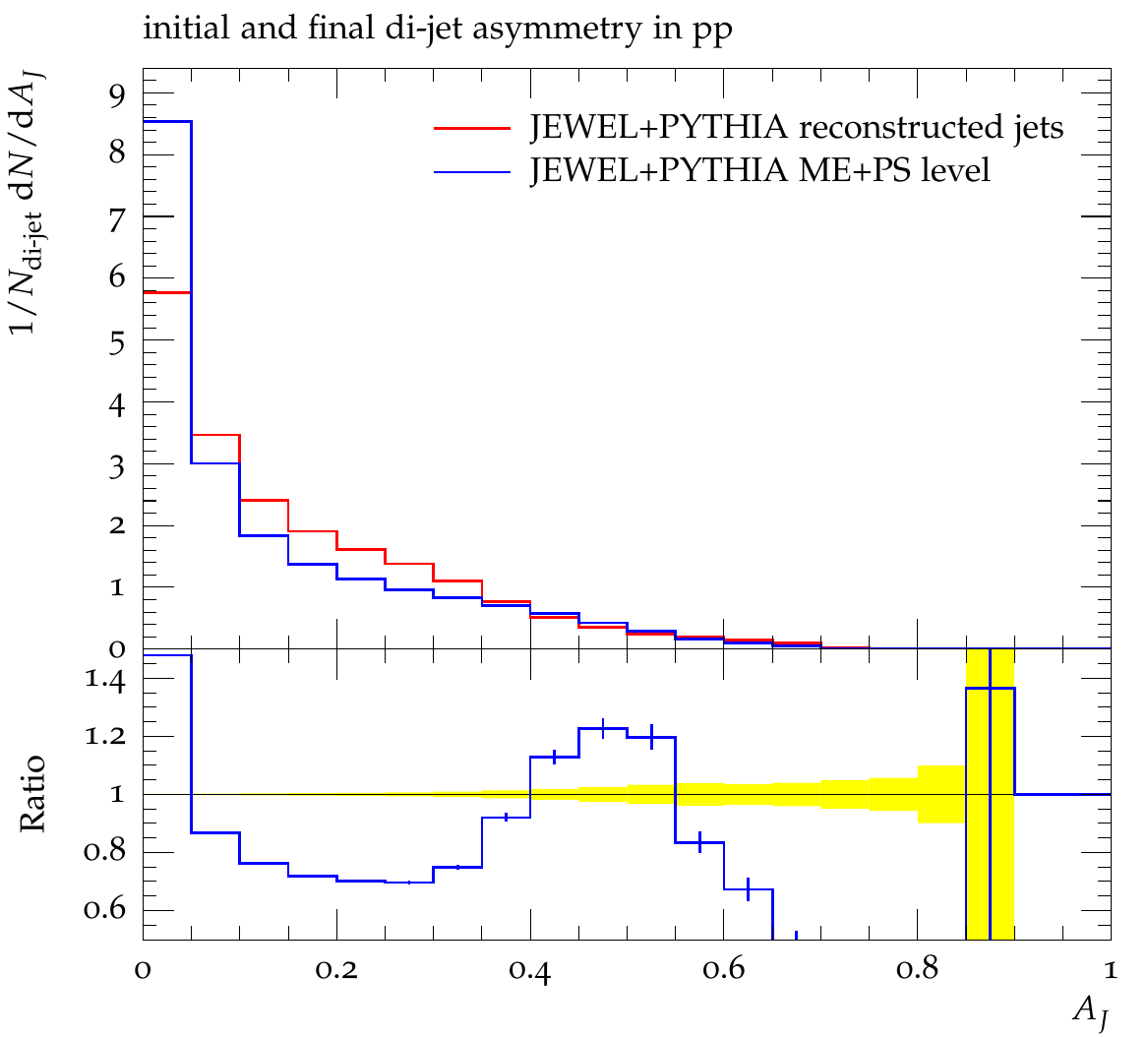}
\caption{Final di-jet asymmetry after jet evolution, hadronisation and jet reconstruction (red) compared to the (partonic) initial asymmetry caused be recoil against initial state and the first final state emission (blue).}
\label{fig::ajin}
\end{figure}


\begin{figure}
	\includegraphics[width=\linewidth]{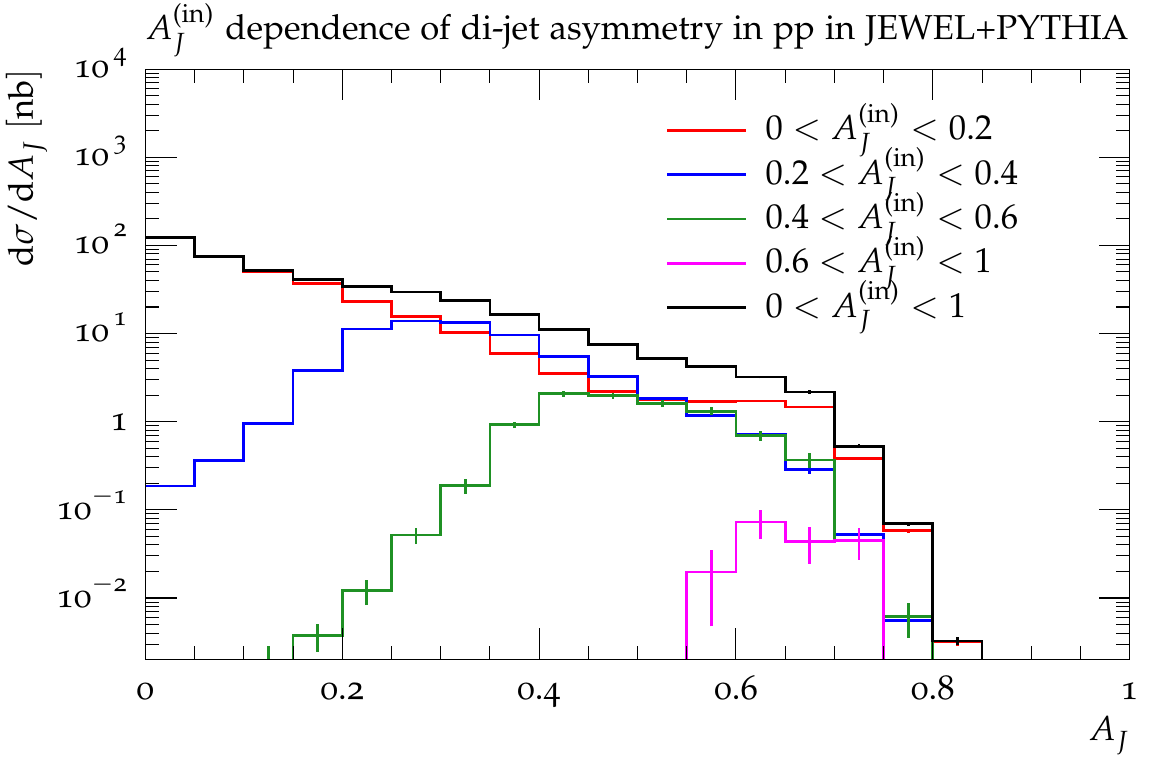}
	\caption{Contributions of the different bins in initial asymmetry $A_J^\text{(in)}$ to the final di-jet asymmetry in p+p events.}
	\label{fig::ajbinnedxsecvac}
\end{figure}

Two effects contribute to the asymmetry in p+p events: recoil against initial and final state emissions and mismatch between the outgoing partons from the matrix element and the reconstructed jets. 
The recoil from each emission in the parton shower has to be compensated within the event. 
Since the incoming partons have to stay parallel to the beams, the recoil from initial state emissions is taken by the final state partons. Similarly, the recoil of the first emission from the final state is transferred to the other final state parton.  The recoil of all later final state emissions can be compensated more locally, i.e.\  by partons originating from the same parent parton and is more likely to end up reconstructed in the same jet. To quantify the effect of the recoil distribution we define an initial configuration which consists of the final partons of the $2\to 2$ configuration and includes the recoil from all initial state and the first final state emission. The asymmetry of this initial configuration is shown in figure~\ref{fig::ajin} compared to the final asymmetry after parton showers, hadronisation and jet reconstruction. The initial asymmetry acounts for most of the observed (final) asymmetry, particularly so for large values of $A_J$.
We have further checked, by considering the asymmetry obtained without initial state radiation, that the effect of recoil against the first final state emission is small. 

\begin{figure}
\includegraphics[width=\linewidth]{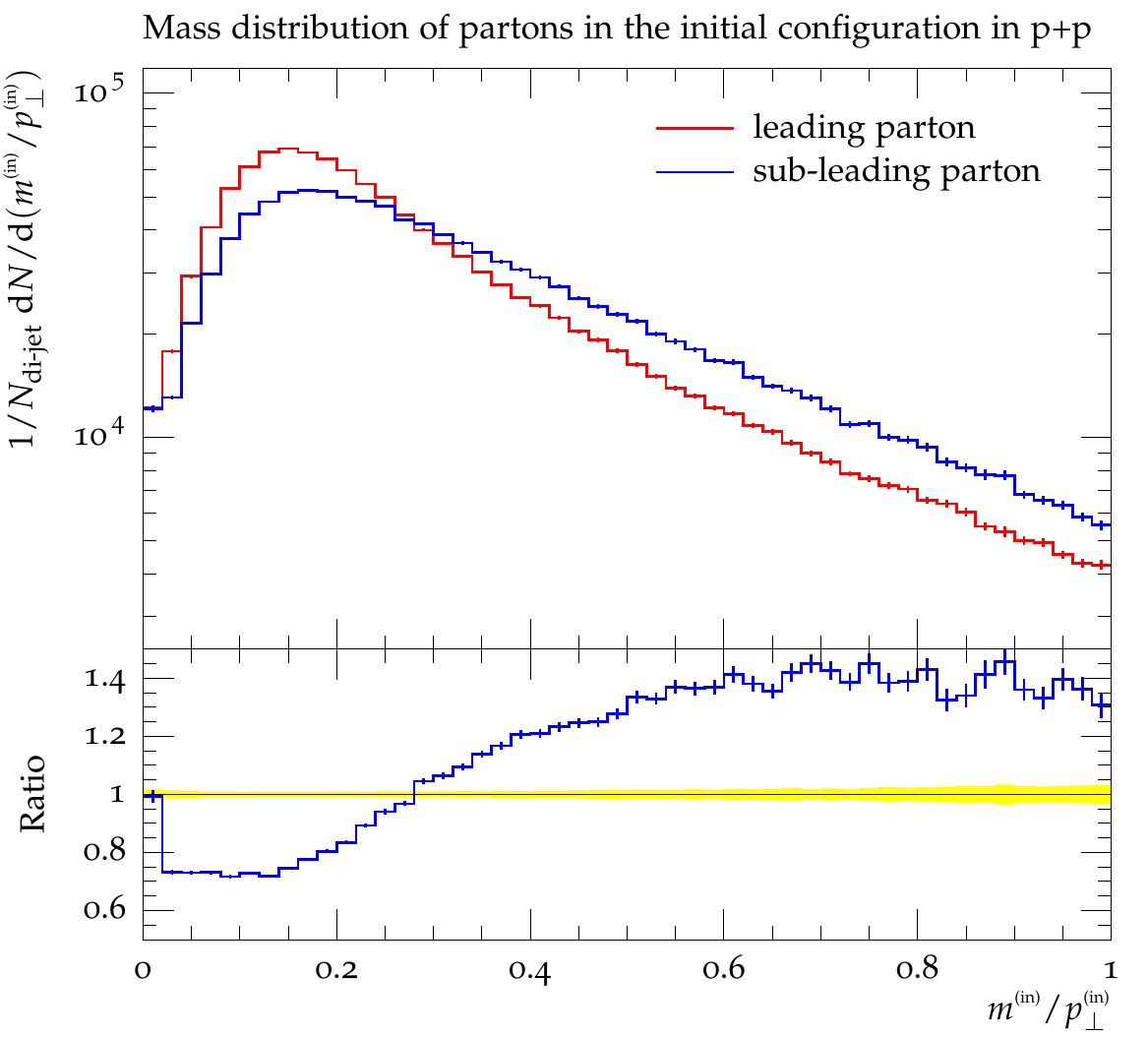}
\caption{Mass to transverse momentum ratio for the partons forming the initial configuration in p+p.}
\label{fig::virtopt}
\end{figure}

Event-by-event correlation between initial and final asymmetry follows from figure~\ref{fig::ajbinnedxsecvac}, where the final asymmetry $A_J$ is shown for different values of the initial asymmetry $A_J^\text{(in)}$. The final asymmetry is indeed not very different from the initial asymmetry, albeit with a clear tendency for the final asymmetry to be larger than the initial one. This increase in asymmetry implies a larger transverse momentum loss for the sub-leading jet. Sub-leading jets originate from initial partons with a higher mass to $\pt$ ratio than leading jets, see figure~\ref{fig::virtopt}. Equivalently, sub-leading jets are those with a softer fragmentation pattern (for the same initial parton $\pt^\text{(in)}$, a larger number of fragments of lower average $\pt$). Thus, the fraction of the initial parton $\pt$ captured within a given reconstruction radius by the jet algorithm is smaller for the sub-leading jet than for the leading jet resulting in an increase of the asymmetry.

\begin{figure}
\includegraphics[width=\linewidth]{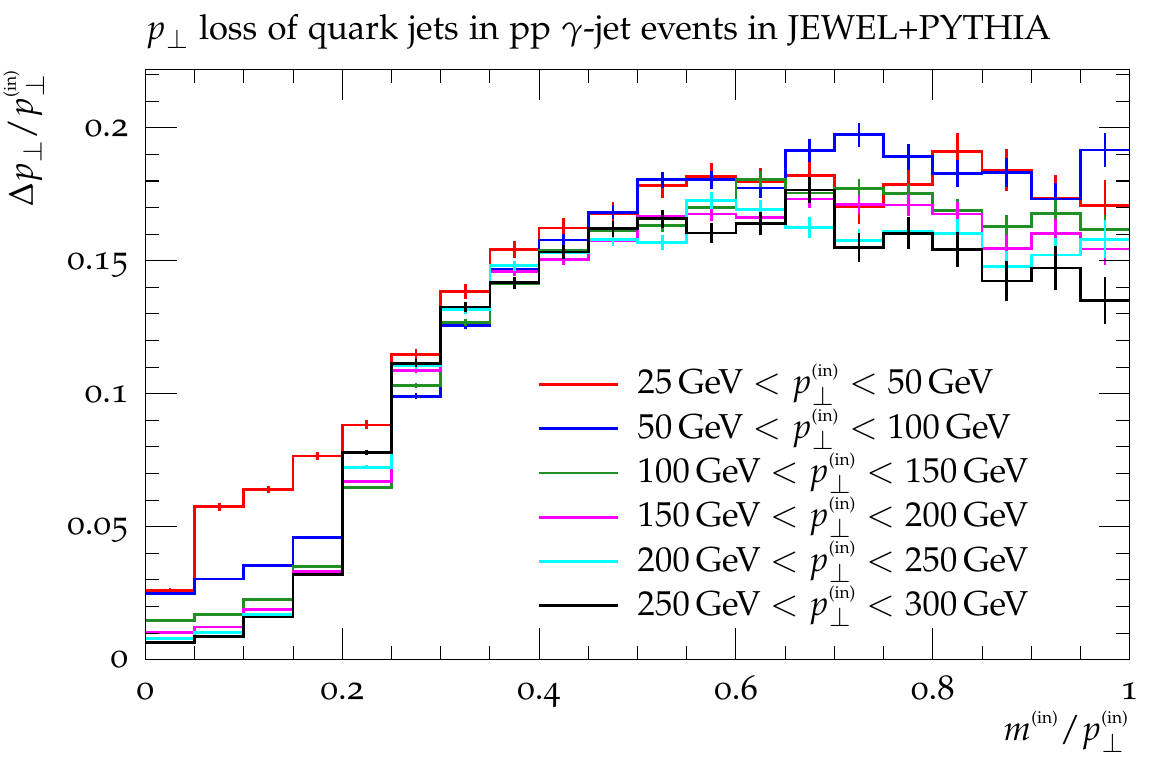}
\caption{Transverse momentum difference between initial parton and reconstructed jet in $\gamma$-jet events in p+p collisions for quark jets reconstructed using the anti-$\kt$ algorithm with $R=0.4$ as function of mass and $\pt$ of the initial parton.}
\label{fig::ptshiftvac}
\end{figure}

The effect of transverse momentum loss when going from initial parton to reconstructed jet can be isolated by considering a sample of $\gamma$-jet events with initial state parton showering disabled.
Here, we associate the hardest final state jet with the initial parton and study the $\pt$ difference between the two. The relative transverse momentum loss $\Delta \pt/\pt^\text{(in)}$ is largely determined by the mass to transverse momentum ratio of the initial parton, as shown in figure~\ref{fig::ptshiftvac}. The $\pt$ loss increases strongly with increasing $m^\text{(in)}/\pt^\text{(in)}$ and then levels off. This saturation occurs because for very large masses increasing the mass further affects mostly the large angle structure already outside the chosen jet reconstruction radius $R$. The point at which the saturation sets in  depends on the reconstruction radius and moves to larger values of $m^\text{(in)}/\pt^\text{(in)}$ with increasing $R$. It should be noted that the $m^\text{(in)}/\pt^\text{(in)}$ distribution is concentrated in the small $m^\text{(in)}/\pt^\text{(in)}$ region which is in the rising part of the $\Delta \pt/\pt^\text{(in)}$ dependence for all reasonable jet radii (cf. figure~\ref{fig::virtopt}). 

\begin{figure}
\includegraphics[width=\linewidth]{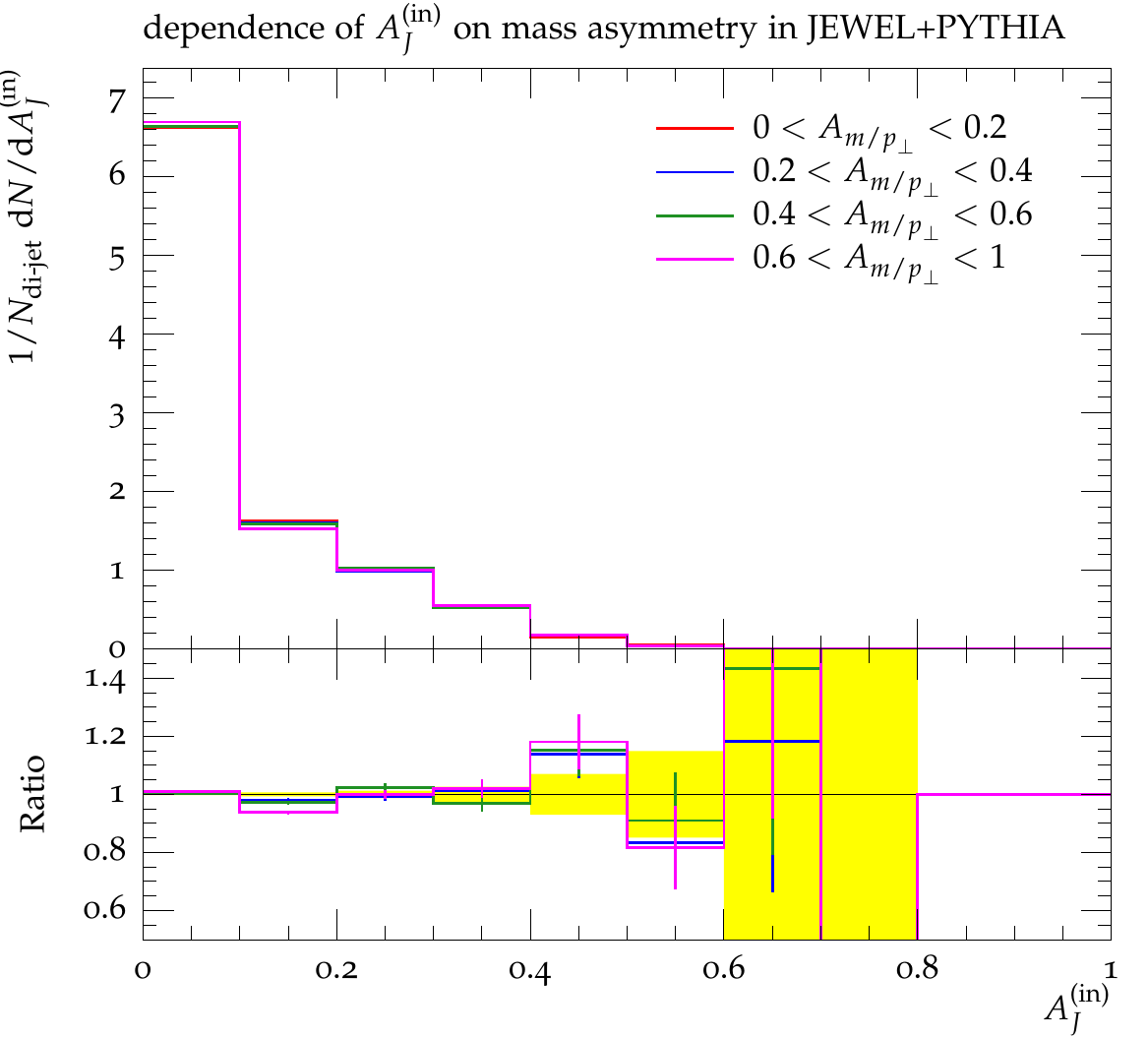}
\caption{Initial di-jet asymmetry binned in mass asymmetry in p+p events.}
\label{fig::ajinmcorr}
\end{figure}

\begin{figure}
\includegraphics[width=\linewidth]{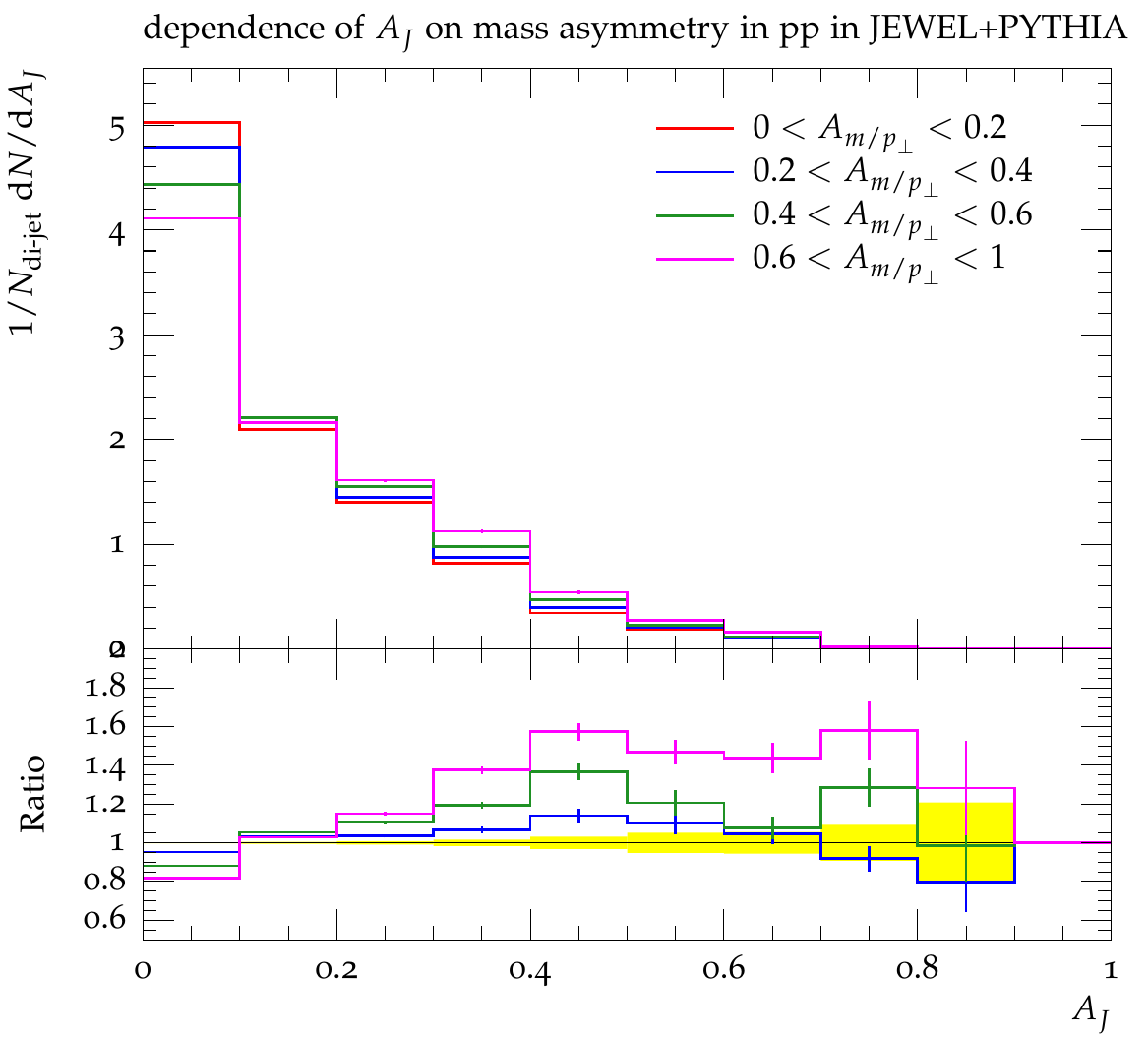}
\caption{Final di-jet asymmetry binned in mass asymmetry in p+p events.}
\label{fig::ajmcorr}
\end{figure}

To ascertain that the mass to transverse momentum ratio of the initial parton is causing the increase in asymmetry when going from initial partons to jets, we 
define, analogously to the $\pt$ asymmetry, an initial mass asymmetry as
\begin{equation}
 A_{m/\pt} = \frac{\left|m^\text{(in)}_1/p^\text{(in)}_{\perp,1} - m^\text{(in)}_2/p^\text{(in)}_{\perp,2}\right|}{m^\text{(in)}_1/p^\text{(in)}_{\perp,1} + m^\text{(in)}_2/p^\text{(in)}_{\perp,2}} \,.
\end{equation}
Figure~\ref{fig::ajmcorr} shows that the final di-jet $\pt$ asymmetry is on average larger in configurations with large mass asymmetry than in those with small mass asymmetry while, figure~\ref{fig::ajinmcorr}, this is clearly not the case for the initial $\pt$ asymmetry.

In summary, the di-jet asymmetry in  p+p collisions results mostly from the recoil against initial state emissions. Although recoil from final state emissions also plays a role, it does so to lesser extent.
This initial asymmetry is increased due to parton showering and jet reconstruction, since the leading jets tend to have a smaller mass and thus fragment harder than softer jets. As the fraction of the parton $\pt$ that ends up in the reconstructed jets depends on the fragmentation pattern, hard jets contain a larger fraction of the initial parton's $\pt$ than soft jets, implying a jet asymmetry larger than the initial parton asymmetry.

\subsection{Di-jet asymmetry in heavy ion collisions}

Having already discarded the in-medium path-length difference between leading and sub-leading jets as a sizeable source of the additional asymmetry observed in heavy ion collisions, we assess now the possible medium effects that could lead to asymmetry increase for both the initial partonic configuration and during further development of the parton shower.

The nuclear modification of parton distribution functions leads to small differences in the initial state evolution. However, its effects were found to be negligible in our simulation. The first final state emission, typically rather hard, occurs on a very short timescale. Even if a medium were to be present at such early times, the point-like spatial scale associated with the hardness of the splitting would render it unresolvable, and thus unaffected, by the medium. Hence, the initial asymmetry $A_J^\text{(in)}$ in heavy ion collisions is unmodified with respect to the proton-proton case. The increase in asymmetry must originate from fluctuations in the vacuum-like fragmentation pattern of the jet and/or of the jet-medium interactions.

An unambigious classification of emissions as vacuum-like or medium induced is not possible and not meaningful. Nevertheless, emissions at scales well above the medium scale cannot be attributed to jet-medium interaction. We refer to this part of the fragmentation pattern as vaccum-like. Indeed, in \textsc{Jewel} the hard fragmentation pattern of the  parton shower is oblivious to medium effects. We believe this to be a correct implementation of jet-medium interaction as hard splittings occur at scales that are well separated from those typical of the medium and, as such, should not be modified by it. 
The vacuum-like fragmentation pattern does, however, play an important role in the medium-induced energy loss. Jets with a softer fragmentation pattern, that is to say with a softer and larger number of constituents, will experience larger loss of $\pt$. As each jet constituent is a candidate for experiencing medium-induced radiation, the larger their number, the larger the medium effect will be. Further, the softer the constituents, the more likely they are to be transported via elastic collisions beyond the reach of the jet reconstruction radius. In this sense, the fluctuations of the jet-medium interaction are amplified in jets with an underlying soft fragmentation pattern.

\begin{figure}
\includegraphics[width=\linewidth]{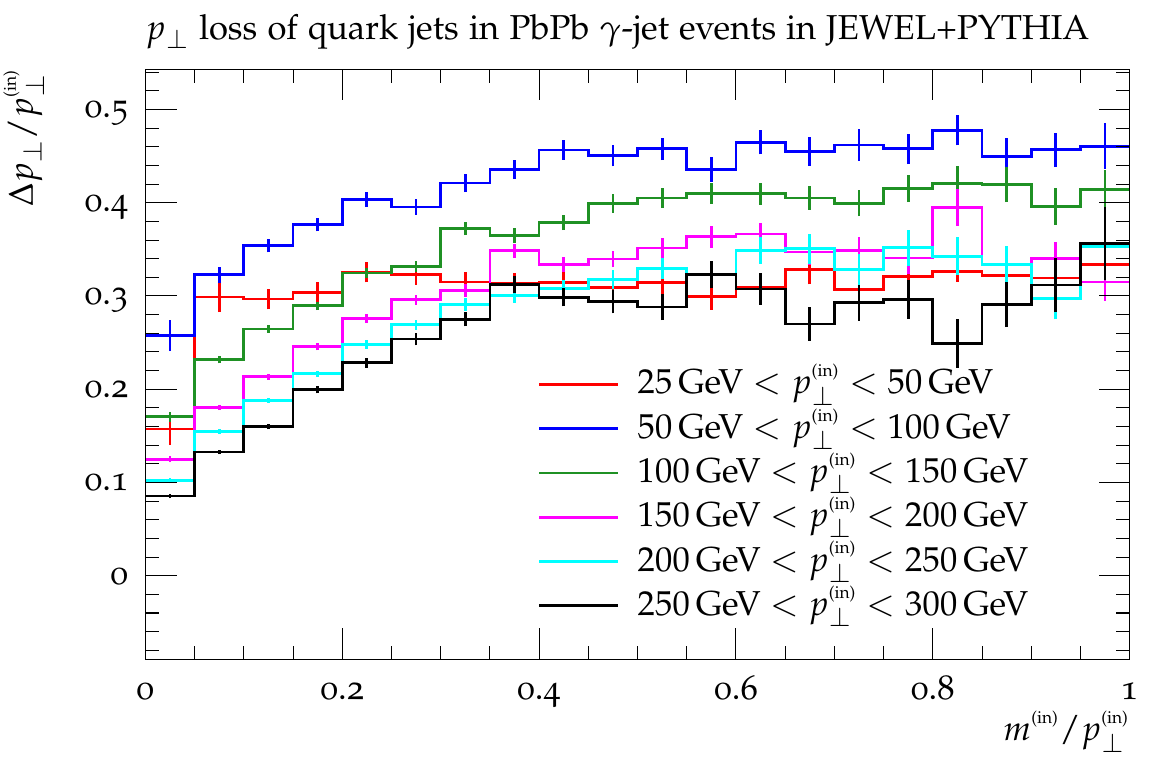}
\caption{Transverse momentum difference between initial parton and reconstructed jet in $\gamma$-jet events in Pb+Pb collisions for quark jets reconstructed using the anti-$\kt$ algorithm with $R=0.4$ as function of mass and $\pt$ of the initial parton.}
\label{fig::ptshiftmed}
\end{figure}

These effects can be seen in the $\pt$ shift shown in Figure~\ref{fig::ptshiftmed}, which was extracted from $\gamma$-jet events in the same way as for p+p collisions. Three observations are in order. First, the transverse momentum loss is larger in Pb+Pb events by nearly a factor of two. Second, the dependence on the $m^\text{(in)}/\pt^\text{(in)}$ ratio is qualitatively similar to the  p+p case and hence the result of vacuum-like dynamics. Finally, this dependence is weaker than in p+p, particularly so for soft jets, showing the importance of medium related fluctuations. This can be clearly seen for the $\unit[25]{GeV} < \pt^\text{(in)} < \unit[50]{GeV}$ bin, where the absence of mass dependence indicates dominance of fluctuations in the medium induced energy loss.

\begin{figure}
\includegraphics[width=\linewidth]{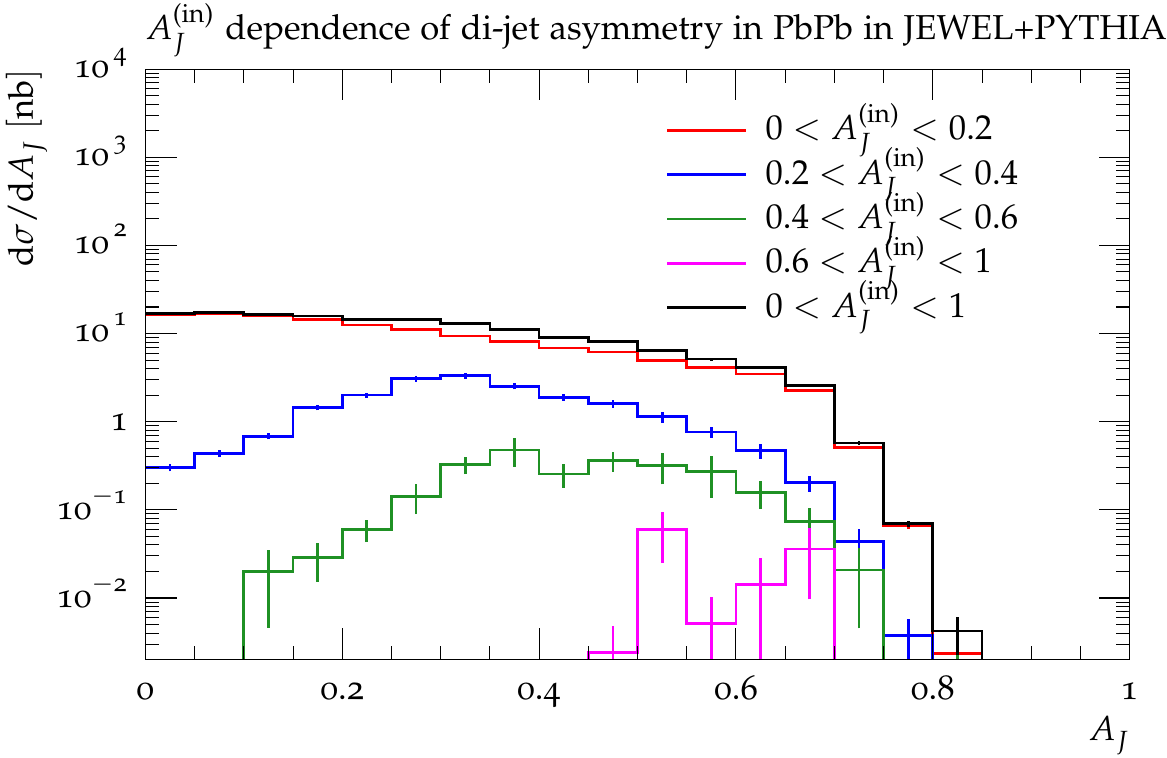}
\caption{Contributions of the different bins in initial asymmetry $A_J^\text{(in)}$ to the final di-jet asymmetry in Pb+Pb events.}
\label{fig::ajbinnedxsecmed}
\end{figure}

\begin{table}
\centering
  \caption{Fraction of di-jets that are lost when going from p+p to Pb+Pb in each $A_J^\text{(in)}$ bin.}
  \label{tab::fractions}
\begin{tabular}{cc}
\hline\noalign{\smallskip}
	$A_J^\text{(in)}$ bin & $(\sigma_\text{pp} - \sigma_\text{PbPb})/\sigma_\text{pp}$ \\
\noalign{\smallskip}\hline\noalign{\smallskip}
	$0.0 < A_J^\text{(in)} < 0.2$ & $0.623 \pm 0.002$ \\
	$0.2 < A_J^\text{(in)} < 0.4$ & $0.699 \pm 0.009$ \\
	$0.4 < A_J^\text{(in)} < 0.6$ & $0.729 \pm 0.034$ \\
	$0.6 < A_J^\text{(in)} < 1.0$ & $0.354 \pm 0.291$ \\
	$0.0 < A_J^\text{(in)} < 1.0$ & $0.637 \pm 0.002$ \\
\noalign{\smallskip}\hline
\end{tabular}
\end{table}

The interplay between vacuum-like and medium related fluctuations can also be seen
in the correlation between initial $A_J^\text{(in)}$ and final  $A_J$ asymmetry for Pb+Pb collisions shown in  figure~\ref{fig::ajbinnedxsecmed}.
Compared to the p+p case, the distributions are broader and, for large initial asymmetries, there is a tendency for the final asymmetry to be smaller. Both features are a direct consequence of medium related fluctuations. Figure~\ref{fig::ajbinnedxsecmed} also reveals that the fractions of di-jets falling into the different $A_J^\text{(in)}$ bins is different from p+p case. The fractions of di-jets that are missing in the Pb+Pb sample compared to p+p are given in table~\ref{tab::fractions}, which shows that the probability for a di-jet to disappear because it fails the cuts increases with initial asymmetry (except for the last bin, which has very poor statistics). This indicates that the systematic increase of initial asymmetry is somewhat larger in heavy ion collisions than in p+p, since on top of the effects already present in the vacuum case jets with a softer fragmentation pattern are also more susceptible to medium modifications.

\begin{figure}
\includegraphics[width=\linewidth]{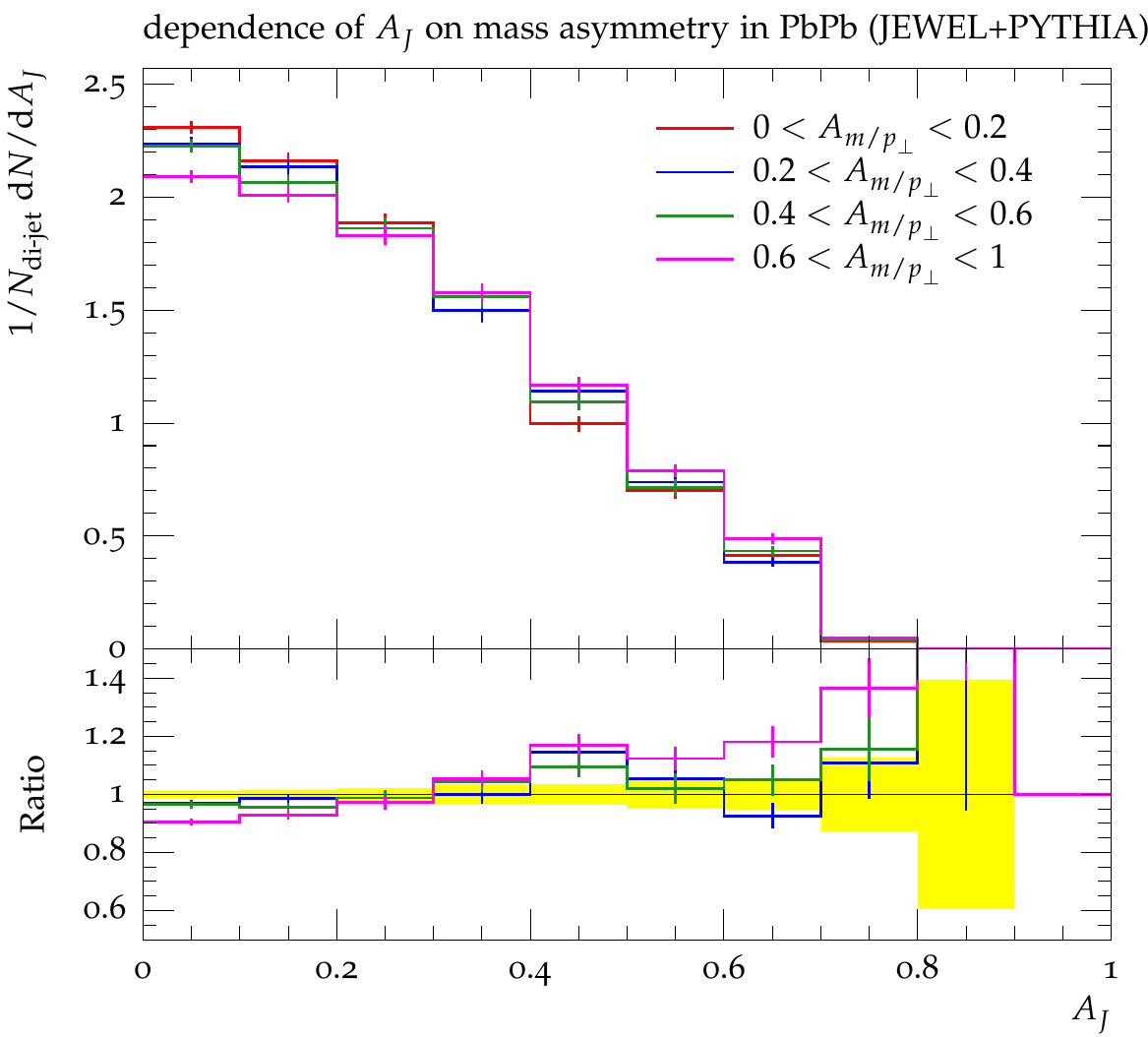}
\caption{Final di-jet asymmetry binned in mass asymmetry in Pb+Pb events.}
\label{fig::ajmcorrmed}
\end{figure}

Finally, figure~\ref{fig::ajmcorrmed} shows the dependence of the final asymmetry on the initial mass asymmetry in Pb+Pb events. The same trend observed in p+p, namely that the final di-jet asymmetry increases with initial mass asymmetry, is seen here highlighting the  importance of fluctuations in the vacuum-like fragmentation in building the observed asymmetry. The dependence is somewhat weaker as result of medium related fluctuations.


\section{Summary \& outlook}
\label{sec:summary}
The main findings of our study can be summarized as follows:

\begin{enumerate}[(i)]
\item the path length difference between the leading and sub-leading jets in a di-jet pair does not play a significant role in generating di-jet asymmetry (momentum imbalance);
\item the increase in di-jet asymmetry in heavy ion collisions is the result of the compound effect of fluctuations in the vacuum-like fragmentation pattern (parton shower features also present in the absence of a medium) and medium related fluctuations;
\item to a large extent, the amount of energy lost from a jet is determined by the mass to transverse momentum ratio of the parton from which it originates.
\end{enumerate}

Although our analysis was carried out in a specific implementation of jet-medium interactions, namely \textsc{Jewel}, it relies on rather generic features and we believe that the main findings should hold in general. The effects leading to these results are properties of the jet evolution in the presence of a medium rather than of the medium itself. It is therefore not surprising that we obtain very similar results with a hydrodynamic background. A source of medium related fluctuations that we did not discuss are the initial conditions of the soft background. The effect of energy loss fluctuations that we observe is therefore probably underestimated. The prominent role played by the hard, i.e.\ vacuum like, fragmentation pattern highlights the importance of a realistic description of the vacuum fragmentation pattern also for jet quenching observables.

Although a number of different models~\cite{Zapp:2013vla,Qin:2010mn,Renk:2012cx,Young:2011qx,Ma:2013pha,Senzel:2013dta,ColemanSmith:2012vr,Casalderrey-Solana:2014bpa,Casalderrey-Solana:2015vaa} have successfully reproduced measurements of the di-jet asymmetry the situation on the theory side has so far been inconclusive. For instance, strong surface bias is found in~\cite{ColemanSmith:2012vr}, while there is hardly any in~\cite{Renk:2012cx}. The authors of~\cite{Ma:2013pha,Senzel:2013dta} notice that the asymmetry of the configuration entering the medium plays a role. However, in these models the parton shower associated to the hard scattering producing the hard partons develops as in vacuum down to the hadronic scale before medium interactions start. Contrary to this ad-hoc factorisation, in \jewel jet evolution and jet-medium interactions happen simultaneously and are dynamically related. The results can therefore not be compared directly. The same is true for~\cite{ColemanSmith:2012vr}, where a dependence on the vacuum fragmentation pattern is observed.

Our finding that the fractional $\pt$ loss depends largely in the initial mass to $\pt$ ratio and only weakly on the initial $\pt$ is in qualitative agreement with the observation that in a strongly coupled scenario the fractional energy loss depends only on the jet opening angle~\cite{Chesler:2015nqz}. The picture of jet quenching that emerges from our study ressonates strongly, and can be seen as strong substantiation for, that put forward in  \cite{CasalderreySolana:2012ef}. There, it was argued that intra-jet coherence properties made the energy loss of jets to be determined by the number of emitters that could be resolved by the medium. Our work establishes that the main driver for determining the number of resolvable emitters is, in fact, the initial mass to transverse momentum ratio (a proxy for the number of vacuum-like splittings in the shower).

The analysis carried out in this short note is one that can be replicated for any observable of interest. As in this study we found that the di-jet asymmetry is sensitive to the fluctuations of the fragmentation pattern, we believe that further analyses can identify observables sensitive to other jet-medium interaction properties.

\begin{acknowledgements}
We are grateful to Liliana Apolin\'ario, N\'estor Armesto, Paul Chesler, Doga Can Gulhan, Yen-Jie Lee, Krishna Rajagopal, Gavin Salam, and Carlos Salgado
for helpful conversations over the course of this work. 
This work was partly supported by Funda\c{c}\~{a}o para a Ci\^{e}ncia e a Tecnologia (Portugal) under project CERN/FIS-NUC/0049/2015, contract `Investigador FCT - Development Grant' (JGM), and postdoctoral fellowship SFRH/BPD/102844/2014 (KCZ).
\end{acknowledgements}

\end{document}